\documentstyle[ichep,12pt]{article}
\title{\normalsize A MORE EFFECTIVE POTENTIAL\thanks{Research
       supported by the U.S. Department of Energy under contract
       DE-FG04-84ER40166.}}
\author{Kevin Cahill\\
        Department of Physics and Astronomy\\
        University of New Mexico\\
        Albuquerque, New Mexico 87131-1156, USA}

\begin{document}
\finalcopy

\maketitle

\abstract{
In theories with spontaneous symmetry breaking,
the conventional effective potential possesses
a defective loop expansion.  For such theories,
the exact effective potential $V(\phi_c,T)$ is real and convex,
but its perturbative series is complex and concave
at small $\phi_c$ and $T$.
A more effective potential
is available.}

\vskip-1pc
\onehead{INTRODUCTION}
The effective potential was introduced
by Goldstone, Salam, and Weinberg$^1$
and by Jona-Lasinio$^2$
in order to discuss broken symmetries.$^3$
Coleman and E.~Weinberg used the effective potential to
show that radiative corrections could break symmetries.$^4$
Linde$^5$ and Weinberg$^6$
later used it to obtain a lower bound on the mass
of the Higgs boson.
West and others have used it
to study the breaking of supersymmetry.$^7$
\par
Bernard,$^{8}$ Weinberg,$^{9}$ and Dolan and Jackiw$^{10}$
extended the effective potential
to finite temperatures
and exhibited the restoration of symmetry at high temperatures.
Much work on the early universe
is based upon the finite-temperature effective potential.
\par
Despite its wide use in particle physics,
the effective potential is flawed.
In theories with spontaneous symmetry breaking,
the loop expansion for the effective potential typically
fails.$^{11}$
In such theories the exact effective
potential is real and convex,$^{12}$
but its perturba-
\break
\newpage
\vglue2pt
\noindent
tive series is neither.
For example, the loop
expansion generated by
the classical potential
$V(\phi) = (\lambda/4)(\phi^2-\sigma^2)^2$
is complex for all temperatures $T$
when $|\phi_c| < \sigma/\sqrt{3}$
and concave (as a function of $\phi_c$) at low temperatures
for $|\phi_c| < \sigma$.
\par
Because of such failures of the loop expansion,
some physicists have turned to nonperturbative
variational$^{13}$ or lattice$^{14}$ techniques.
Yet a perturbative effective potential
would be useful because
it could be readily applied to a wide variety
of theories.
\par
In realistic theories with
spontaneous symmetry breaking, it is only
the contribution of the scalar fields,
not that of the fermions or gauge bosons,
that becomes complex.
Complex values arise
in the loop expansion because
it is not possible to quantize the scalar fields
where the matrix
of second derivatives of the classical potential $V(\phi)$
has negative eigenvalues.
The addition of the linear
perturbation $j\phi$ to the potential $V(\phi)$
shifts its stationary points
but does not change their curvature.
\par
The present paper will show that
by using a quadratic perturbation $j(\phi + a \phi^2)$
one may construct a related, more effective potential that
possesses a real loop expansion.
\goodbreak

\onehead{THE EFFECTIVE POTENTIAL}

For a scalar field $\phi$ described by a hamiltonian $H$,
the Helmholtz free-energy density $A(j,T)$ is defined by
\begin{equation}
e^{-\beta L^3 A(j,T)} =
{\rm Tr} e^{-\beta \left(H + j\!\int\!  \phi(x)\,d^3x\right)}
\end{equation}
where $j$ is a perturbing external current
and $L^3$ is the volume of quantization.
The mean value of the field $\phi$
\begin{equation}
\phi_c(j,T) \equiv \langle \phi \rangle_j = {{\rm Tr} \phi(x)
e^{-\beta \left(H + j\int\!\! \phi(x) d^3x \right)} \over
 {\rm Tr} e^{-\beta \left(H + j\int\!\! \phi(x) d^3x\right)}}
\end{equation}
is a derivative of the Helmholtz potential
\begin{equation}
\phi_c(j,T) = {\partial A(j,T) \over \partial j}.   \label{phisdA}
\end{equation}
The finite-temperature effective potential $V(\phi_c,T)$
is defined $^{8-10}$
as a Legendre transform of the Helmholtz potential
\begin{equation}
V(\phi_c,T) \equiv A - j{\partial A \over \partial j}
= A - j \phi_c                                \label{Vdef}
\end{equation}
written as a function of the ``classical field'' $\phi_c$
rather than of the external current:
$V(\phi_c,T)=A(j(\phi_c,T),T) - j(\phi_c,T)\phi_c$.
\par
At its stationary points, the effective potential
describes the unperturbed system.
For from eqs.(\ref{phisdA}) and (\ref{Vdef}), it follows that
the derivative of the effective potential
with respect to the classical field $\phi_c$
is proportional to the external current $j$
\begin{equation}
{\partial V(\phi_c,T) \over \partial \phi_c}
= {\partial j \over \partial \phi_c}
\left( {\partial A \over \partial j}
- \phi_c \right) - j = - j.                        \label{dV}
\end{equation}
Thus at the minima of $V(\phi_c,T)$
the current $j$ must vanish,
$j = - \partial V(\phi_c,T) / \partial \phi_c = 0$.
At zero temperature the minimum of the effective
potential is the energy density of the vacuum.
\par
The factor $\exp(-\beta j \int \phi(x) d^3x)$
ensures that
the derivatives $\partial \phi_c / \partial j$
and $\partial j / \partial \phi_c$ are negative or zero.
So by differentiating the formula (\ref{dV})
with respect to $\phi_c$,
we see that the effective potential
is a convex function of $\phi_c$:$^{12}$
\begin{equation}
{\partial^2 V(\phi_c,T) \over \partial \phi_c^2}
= - {\partial j  \over \partial \phi_c} \ge 0.
\end{equation}
\goodbreak

\twohead{The Exact Effective Potential for a Free Field}

For the free scalar field of mass $m$,
the Helmholtz potential is
\begin{equation}
A(j,T) = -{\ln Z(\beta) \over \beta L^3} - {j^2 \over 2m^2}
\end{equation}
where $Z(\beta)= {\rm Tr} \exp(-\beta H)$ is the partition function
with logarithm
\begin{equation}
-\ln Z(\beta) =
\int\!\! {\beta \omega_k \over 2}
+ \ln\left(1 - e^{-\beta \omega_k }\right)
{L^3 d^3k \over (2\pi)^3}.  \label{Z}
\end{equation}
\par
The mean value $\langle\phi\rangle$ is
$\phi_c = \partial A(j,T)/ \partial j = - j/m^2$.
So the exact, finite-temperature
effective potential for this system is
\begin{eqnarray}
V(\phi_c,T) & = & A(j,T) - j \phi_c \qquad\qquad  \nonumber \\
& = & {\textstyle { 1 \over 2}} m^2 \phi_c^2 \label{Vexact} \\
& & \mbox{} + \int\!\! {\omega_k \over 2}
+ {\ln\left(1 - e^{-\beta \omega_k }\right) \over \beta}
{d^3k \over (2\pi)^3}. \nonumber
\end{eqnarray}
The effective potential $V(\phi_c,T)$
is real and convex.  At its stationary point
$\phi_c=0$, the external current $j = - m^2\phi_c$ vanishes.
\goodbreak

\onehead{THE ONE-LOOP EFFECTIVE POTENTIAL}

\par
For a scalar field $\phi$
with potential $V(\phi)$,
we may obtain the one-loop approximation to the Helmholtz
potential by replacing the altered potential
$V_j(\phi) = V(\phi) + j \phi$
by its quadratic expansion
\begin{equation}
V_j(\phi) \approx V_j(\phi_0)
+ {\textstyle { 1 \over 2}}
{\partial^2 V_j(\phi_0) \over \partial \phi^2}
(\phi - \phi_0)^2
\end{equation}
about its minimum $\phi_0$ which is a root of the equation
$V_j^\prime(\phi) = V^\prime(\phi) + j = 0$.
The mean value $\phi_c$ is now $\phi_0$.
Thus $V_j^{\prime\prime}(\phi_0) = V^{\prime\prime}(\phi_c)$,
and so
\begin{equation}
V_j(\phi) \approx V(\phi_c) + j\phi_c
+ {\textstyle { 1 \over 2}}
{\partial^2 V(\phi_c) \over \partial \phi^2}
(\phi - \phi_c)^2.
\end{equation}
\par
In this approximation
the Helmholtz potential is effectively that of a free scalar field
of mass $m = \sqrt{ V_j^{\prime\prime}(\phi_0)}$.
So by using our exact result (\ref{Vexact})
for that system, we find
\begin{eqnarray}
e^{-\beta L^3 A(j,T)} & = &
{\rm Tr} e^{-\beta \left(H + j\int\!\! \phi(x) d^3x\right)}
\nonumber \\
& \approx & e^{-\beta L^3 \left( V(\phi_c) + j\phi_c \right)} Z(\beta),
\end{eqnarray}
where $Z(\beta)$ is the partition function (\ref{Z})
for a free field of mass $m = \sqrt{ V^{\prime\prime}(\phi_c) } $.
Thus at the one-loop level,
the Helmholtz potential is
\begin{equation}
A_1(j,T) = V(\phi_c) + j\phi_c - {\ln Z(\beta) \over \beta L^3};
\end{equation}
and so the effective potential is
\begin{eqnarray}
V_1(\phi_c,T) & = & V(\phi_c) \\
& & \mbox{} + \int\!\! {\omega_k \over 2}
+ {\ln\left(1 - e^{-\beta \omega_k }\right) \over \beta}
{d^3k \over (2\pi)^3} \nonumber
\end{eqnarray}
with
$\omega_k = \sqrt{ k^2 + V^{\prime\prime}(\phi_c)}$.
When $V^{\prime\prime}(\phi_c) < 0$,
the frequency $\omega_k$ becomes
complex for small enough $k$,
and the loop expansion for
the effective potential fails.
\par
For the potential
$V(\phi) = \lambda \left( \phi^2 - \sigma^2 \right)^2/4$,
the second derivative
$V^{\prime\prime}(\phi) = \lambda \left( 3\phi^2 - \sigma^2 \right)$
is positive only for fields $|\phi|$
greater than
$\sigma / \sqrt{3}$.
It is not possible to quantize the
approximate, altered theory
about smaller values of $\phi$.
\par
With minimal renormalization
the one-loop effective potential is
\begin{eqnarray}
\lefteqn{V_1(\phi_c,T) = {\lambda\over4}
(\phi_c^2-\sigma^2)^2}\nonumber\\
& & \mbox{} + {\lambda^2 (3\phi_c^2 - \sigma^2 )^2 \over 64\pi^2}
\left[ \ln\lambda (3\phi_c^2 - \sigma^2 )
+ {\textstyle { 1 \over 2}} \right] \nonumber \\
& & \mbox{} + {T^4\over2\pi^2}\int_0^\infty x^2
\ln\left( 1 - e^{-\rho(x)} \right) dx \label{olep}
\end{eqnarray}
where $\rho(x)$ is
$\rho(x) = \sqrt{x^2 + \lambda(3\phi_c^2-\sigma^2)/T^2}$.
To this expression one may add arbitrary, finite
contributions like
$\lambda^2(A\phi_c^4+B\phi_c^2+C)$
from the renormalization of the
hamiltonian $H$.
Since the effective potential is
a sum of a function of $\phi^2 - \sigma^2$
and a function of $3\phi_c^2 - \sigma^2$,
it is obviously not convex.
With minimal renormalization $V_1(\phi_c,T)$
is concave for $|\phi| < \sigma$ for small
$\lambda$ at low temperatures $T$.
\goodbreak

\onehead{A MORE EFFECTIVE POTENTIAL}

\par
We may change the curvature of $V(\phi)$ by defining
a more general Helmholtz potential $A(j,T;P)$
in which the linear probe $j \phi$ is replaced by
a quadratic polynomial $j P(\phi)$
\begin{equation}
e^{-\beta L^3 A(j,T;P)} =
{\rm Tr} e^{-\beta \left(H + \int\!\! j P(\phi) d^3x \right)}.
\label{meadef}
\end{equation}
On the one hand,
it is clear that by this device
we have not introduced any new divergences
into the theory.
On the other hand,
it is also clear that the polynomial $P(\phi)$
is singular itself and requires renormalization.
Now the derivative of $A(j,T;P)$
with respect to the external current $j$
\begin{equation}
{\partial A(j,T;P) \over \partial j} = P_c
\end{equation}
is the mean value of the polynomial $P(\phi)$
\begin{equation}
P_c = \langle P(\phi) \rangle = {{\rm Tr} P(\phi(x))
e^{-\beta \left(H + \int\!\! j P(\phi) d^3x \right)} \over
{\rm Tr} e^{-\beta \left(H + \int\!\! j P(\phi) d^3x \right)}}.
\end{equation}
The classical field $\phi_c$ is still the mean value
$\langle \phi \rangle$
\begin{equation}
\phi_c = \langle \phi \rangle =  {{\rm Tr} \phi(x)
e^{-\beta \left(H + \int\!\! j P(\phi) d^3x \right)} \over
{\rm Tr} e^{-\beta \left(H + \int\!\! j P(\phi) d^3x \right)}}.
\end{equation}

\par
We may now define the more effective potential
$V(\phi_c, T; P)$ as the Legendre transform
\begin{eqnarray}
V(\phi_c, T; P) & = &
A(j,T;P) - j {\partial A(j,T;P) \over \partial j}\nonumber \\
& = & A(j,T;P) - j P_c. \label{mepdef}
\end{eqnarray}
Like the effective potential,
the more effective potential
at its minima
describes the unperturbed system.
For where $V(\phi_c, T; P)$ is stationary,
the external current $j$ vanishes
\begin{eqnarray}
0 & = & {\partial V(\phi_c,T; P)
\over \partial \phi_c}\nonumber \\
& = & {\partial j \over \partial \phi_c}
\left( {\partial A(j,T;P) \over \partial j}
- P_c \right)
- j {\partial P_c \over \partial \phi_c} \nonumber \\
& = & - j {\partial P_c \over \partial \phi_c} \label{j=0}
\end{eqnarray}
unless exceptionally $\partial P_c/\partial \phi_c = 0$.
The more effective potential is real
but not necessarily convex.
\goodbreak

\onehead{WHAT EFFECTIVE POTENTIALS ARE}

\par
The meaning of an effective potential is clearest
in the zero-temperature limit $\beta \to \infty$.
At $T=0$ the Helmholtz potential
is the energy density
$A(j,0;P) = E_j/L^3$ of the
eigenstate $|j\rangle$ of the
altered hamiltonian
\begin{equation}
\left( H + \int j P(\phi) d^3x \right)
| j \rangle = E_j | j \rangle
\end{equation}
with minimum energy $E_j$.
Thus the effective potential $V(\phi_c,0;P)=A(j,0;P)-jP_c$ is
the mean value of the hamiltonian density in the state $|j\rangle$:
\begin{equation}V(\phi_c,0;P) = {\langle j|H|j\rangle \over L^3} .
\end{equation}
By eq.(\ref{j=0}),
the effective potential $V(\phi_c,0;P)$ at its minimum is the
energy density of the ground state $|0\rangle$
of the unperturbed theory, {\it i.e.,} the energy density
of the physical vacuum.
\par
By differentiating its definition (\ref{meadef}),
we may find that
the general Helmholtz potential $A(j,T;P)$ at finite temperatures
is the mean value of the
altered hamiltonian density $\left[H + \int j P(\phi) d^3x\right]/L^3$
in the mixture
\begin{equation}
\rho = {e^{-\beta \left[H + \int j P(\phi) d^3x\right] }
\over {\rm Tr}
e^{-\beta \left[H + \int j P(\phi) d^3x\right] }}.
\end{equation}
The finite-temperature effective potential $V(\phi_c,T;P)$ is
the mean value of the hamiltonian density $H/L^3$ in this mixture.
\goodbreak

\onehead{AN EXAMPLE}

\par
Let us construct
the more effective potential
$V(\phi_c,T;P)$ for the classical potential
$V(\phi) = (\lambda /4) \left( \phi^2 - \sigma^2 \right)^2$
with polynomial $P(\phi) = \phi^2/2$.
We shall see that $V(\phi_c,T;P)$ is real.
\par
The stationary points $\phi_0$ of the altered potential
$V_j(\phi) = V(\phi) + j\phi^2 / 2$
are $\phi_{0\pm} = \pm \sqrt{\sigma^2 - j/\lambda}$.
Let us quantize about the positive root
$\phi_c = \phi_{0+}$.
The mass squared is now
$m^2 = V_j^{\prime\prime}(\phi_c) = 2\lambda\phi_c^2$
which is never negative.
Since the current $j = \lambda(\sigma^2 - \phi^2_c)$,
this mass squared is also
$m^2 = 2\lambda\sigma^2 - 2j$.
\par
At the one-loop level, the Helmholtz free-energy density $A(j,T;P)$ is
\begin{equation}A_1(j,T;P) = V(\phi_c) + j{\phi_c^2\over2}
- {\ln Z(\beta) \over \beta L^3}
\end{equation}
in which $\phi_c$ is to be expressed in terms of
the current $j$ and $Z(\beta)$ is
the partition function (\ref{Z})
for $m^2 = 2\lambda\sigma^2 - 2j$.
Thus the Helmholtz potential is
\begin{eqnarray}
\lefteqn{A_1(j,T;P) = {j\sigma^2\over2} - {j^2\over4\lambda}} \\
& & \mbox{} + \int\!\! \left[ {\omega_k \over 2}
+ {\ln\left(1 - e^{-\beta \omega_k }\right) \over \beta} \right]
{d^3k \over (2\pi)^3} \nonumber
\end{eqnarray}
with
$\omega_k = \sqrt{ k^2 + 2\lambda\sigma^2 - 2j}$.
\par
According to its definition (\ref{mepdef}),
the one-loop effective potential $V_1(\phi_c,T;P)$ is
\begin{equation}
V_1(\phi_c, T; P)
= A_1(j,T;P) - j {\partial A_1(j,T;P) \over \partial j}.
\end{equation}
By performing the indicated differentiation with respect to $j$,
we may express $V_1(\phi_c,T;P)$
in terms of the classical field $\phi_c$ as
\begin{eqnarray}
\lefteqn{V_1(\phi_c,T;P) = {\lambda\over4} (\phi_c^2-\sigma^2)^2}
\nonumber \\
& & \mbox{} + \int {d^3k \over (2\pi)^3} \left[{\omega_k \over 2}
+ { \lambda(\sigma^2-\phi_c^2) \over 2 \omega_k} \right. \nonumber \\
& & \mbox{} + \left. {\ln\left(1 - e^{-\beta \omega_k }\right) \over \beta}
+ {\lambda(\sigma^2-\phi_c^2)
\over \omega_k \left(e^{\beta \omega_k } - 1 \right) } \right]
\end{eqnarray}
with $\omega_k = \sqrt{ k^2 + 2\lambda\phi^2 }$.
The more effective potential is real for
all $\phi_c$ and all $T$.
The first two terms in the integral
require renormalization; the second two
are finite and vanish at $T=0$.
\par
After renormalizing both $H$ and $P(\phi)$,
using a momentum cutoff, I found for the zero-temperature
more effective potential the expression
\begin{eqnarray}
\lefteqn{V_1(\phi_c,0;P) = {\lambda\over4} (\phi_c^2-\sigma^2)^2 }
\label{mepT=0} \\
& & \mbox{} + {\lambda^2 \phi_c^2 \over 32\pi^2}
\left[ (4 \sigma^2 - 2 \phi^2) \ln2\lambda\phi_c^2
+ 4 \sigma^2 - 3 \phi^2 \right] \nonumber
\end{eqnarray}
apart from finite counterterms.
To this expression one must add the same
arbitrary, finite contributions
$\lambda^2(A\phi_c^4 + B\phi_c^2 + C)$
from the renormalization of the hamiltonian $H$
that one adds to the ordinary effective potential (\ref{olep}),
and one may add an arbitrary, finite term of the form
$\lambda^2 D (\sigma^2 - \phi_c^2)^2$
from the renormalization of the perturbation $j\phi^2/2$.
\par
At finite temperatures the more effective potential is
the sum of this zero-temperature potential
and a temperature-dependent term
that arises from the last two terms in the preceding integral:
\begin{eqnarray}
\lefteqn{V_1(\phi_c,T;P) = V_1(\phi_c,0;P)} \nonumber \\
& & \mbox{} + {T^4\over2\pi^2}\int_0^\infty \!\! x^2 \left[
\ln\left( 1 - e^{-r(x)} \right) \right. \nonumber \\
& & \mbox{} + \left. {\lambda(\sigma^2 - \phi_c^2)
\over T^2 r(x) \left( e^{r(x)} - 1 \right) } \right] \! dx.
\label{mepT>0}
\end{eqnarray}
where $r(x)$ is
$r(x) = \sqrt{x^2 + (2\lambda\phi_c^2/T^2)}$.
\par
The formula (\ref{mepT=0})
for the more effective potential is not bounded below.
As $|\phi_c| \to \infty$, the term proportional to
$-\phi_c^4\ln\phi_c^2$ dominates.
Thus just as the usual effective potential with
a linear perturbation $j\phi$ is inappropriate for
$|\phi_c| < \sigma$, so too the more effective potential
gradually loses its appropriateness as
$|\phi_c|$ exceeds $\sigma$.
So in this model,
one may use the more effective potential $V(\phi_c,T;P)$
for $|\phi_c| < \sigma$ and the usual effective potential (\ref{olep})
for $|\phi_c|>\sigma$.
At all temperatures,
the two effective potentials match at $|\phi_c| = \sigma$,
but their derivatives differ.
\par
For this model,
the finite-temperature more effective potential (\ref{mepT>0})
does not have a minimum at $\phi_c =0$ even at very high temperatures.
To the extent that this result transfers to
more complicated models, the inference is that it is
the fermions and gauge bosons but not the Higgs bosons
that restore symmetry at high temperatures.
\par
I have benefited from conversations
with D.~Bailin, I.~Hinchliffe, J.~Jers\'{a}k,
S.~Mandelstam, R.~Reeder, H.~Stapp,
and P.~Stevenson.

\goodbreak

\end{document}